\documentclass[10pt]{bmc_article}    

\usepackage{cite} % Make references as [1-4], not [1,2,3,4]
\usepackage{url}  % Formatting web addresses  
\usepackage{ifthen}  % Conditional 
\usepackage{multicol}   %Columns
\usepackage{multirow}
\usepackage[utf8]{inputenc} %unicode support
\usepackage{graphicx}
\usepackage{amsmath}

\newboolean{publ}

\newenvironment{bmcformat}
{\begin{raggedright}\baselineskip20pt\sloppy\setboolean{publ}{false}}
  {\end{raggedright}\baselineskip20pt\sloppy}

\begin{document}
\begin{bmcformat}
  
  
  \title{The geography and carbon footprint of mobile phone use in 
    Cote d'Ivoire}
  
  %%%%%%%%%%%%%%%%%%%%%%%%%%%%%%%%%%%%%%%%%%%%%% 
  %%                                          %% 
  %% Enter the authors here                   %%
  %%                                          %% 
  %% Ensure \and is entered between all but   %%
  %% the last two authors. This will be       %%
  %% replaced by a comma in the final article %%
  %%                                          %% 
  %% Ensure there are no trailing spaces at   %% 
  %% the ends of the lines                    %%     	
  %%                                          %% 
  %%%%%%%%%%%%%%%%%%%%%%%%%%%%%%%%%%%%%%%%%%%%%% 

  \author{Vsevolod Salnikov$^1$%
    \email{Vsevolod Salnikov - vs.salnikov@gmail.com}%
    \and
    Daniel Schien$^2$%
    \email{Daniel Schien - daniel.schien@bristol.ac.uk}%
    \and 
    Hyejin Youn$^3$%
    \email{Hyejin Youn - visang@santafe.edu}%
    \and
    Renaud Lambiotte$^1$%
    \email{Renaud Lambiotte - renaud.lambiotte@fundp.ac.be}
    and
    Michael T.~Gastner\correspondingauthor$^{4,5}$%
    \email{Michael T.~Gastner\correspondingauthor - m.gastner@bristol.ac.uk}
  }
  
  %%%%%%%%%%%%%%%%%%%%%%%%%%%%%%%%%%%%%%%%%%%%%% 
  %%                                          %% 
  %% Enter the authors' addresses here        %%
  %%                                          %% 
  %%%%%%%%%%%%%%%%%%%%%%%%%%%%%%%%%%%%%%%%%%%%%% 
  
  \address{%
    \iid(1)naXys, University of Namur, Rempart de la Vierge 8, 5000 
    Namur, Belgium\\
    \iid(2)Department of Computer Science, University of Bristol,
    Merchant Venturers Building, Woodland Road, Bristol BS8 1UB, UK\\
    \iid(3)Santa Fe Institute, 1399 Hyde Park Road, Santa Fe, New
    Mexico 87501, USA\\
    \iid(4)Department of Engineering Mathematics, University of Bristol,
    Merchant Venturers Building, Woodland Road, Bristol BS8 1UB, UK\\
    \iid(5)Institute of Technical Physics and Materials Science,
    Research Centre for Natural Sciences, Hungarian Academy of
    Sciences, P.O. Box 49, H-1525 Budapest, Hungary
}%

\maketitle

%%%%%%%%%%%%%%%%%%%%%%%%%%%%%%%%%%%%%%%%%%%%%% 
%%                                          %% 
%% The Abstract begins here                 %%
%%                                          %% 
%%%%%%%%%%%%%%%%%%%%%%%%%%%%%%%%%%%%%%%%%%%%%% 

\begin{abstract}
  The newly released Orange D4D mobile phone data base provides new
  insights into the use of mobile technology in a developing country.
  Here we perform a series of spatial data analyses that reveal
  important geographic aspects of mobile phone use in Cote d'Ivoire.
  We first map the locations of base stations with respect to the
  population distribution and the number and duration of calls at each
  base station.
  On this basis, we estimate the energy consumed by the mobile phone
  network.
  Finally, we perform an analysis of inter-city mobility, and identify
  high-traffic roads in the country. 
\end{abstract}

\ifthenelse{\boolean{publ}}{\begin{multicols}{2}}{}
  
  %%%%%%%%%%%%%%%%%%%%%%%%%%%%%%%%%%%%%%%%%%%%%% 
  %%                                          %% 
  %% The Main Body begins here                %%
  %%                                          %% 
  %% use \cite{...} to cite references        %%
  %% \cite{koon} and                         %%
  %% \cite{oreg,khar,zvai,xjon,schn,pond}    %%
  %% \nocite{smith,marg,hunn,advi,koha,mouse}%%
  %%                                          %% 
  %%%%%%%%%%%%%%%%%%%%%%%%%%%%%%%%%%%%%%%%%%%%%% 
  
  %%%%%%%%%%%%%% 
  %% Keywords %%
  \section*{Keywords}
  mobile phone, Cote d'Ivoire, carbon footprint, human mobility
  
  %%%%%%%%%%%%%%%%%% 
  %% Introduction %% 
  \section{Introduction}
  The availability of mobile phone records has revolutionised our
  ability to perform large-scale studies of social networks and human
  mobility.
  Traditionally, researchers had to rely on a combination of surveys,
  census data and vehicle counting.
  These methods are costly and time consuming so that data were
  collected either infrequently or for small population samples only.
  In the last few years, while searching for innovative methods to
  circumvent these limitations, researchers have turned their attention
  to mobile phones as sensors to collect communication and mobility
  data~\cite{Becker_etal13}.
  The vast majority of studies were carried out in developed countries
  where mobile communication competes with established landline
  technologies.
  However, mobile phones are nowadays commonplace in developing
  countries too.
  Especially in Africa, mobile phones now provide affordable
  telecommunication where no alternative had previously
  existed~\cite{HeeksJagun07,Singh09}. 
  
  The data bases for Cote d'Ivoire, made accessible during the Orange
  D4D challenge~\cite{Blondel_etal12}, present the first opportunity to
  analyse a nationwide mobile phone network in Africa.
  The data are obtained from the so-called Call Detail Records (CDRs)
  which contain an approximate location of mobile phones every time they
  connect to a cell tower (e.g.\ due to a phone call).
  A growing body of research has shown that CDRs can accurately
  characterise many aspects of human mobility~\cite{Hoteit_etal13}.
  Practical examples include the tracking of population displacements
  after disasters~\cite{Bengtsson_etal11,Lu_etal12}, the estimation of
  traffic volumes in cities~\cite{Wang_etal12}, the calculation of
  carbon emissions due to commuting~\cite{Isaacman_etal11} and transport
  mode inference~\cite{Wang_etal10}.
  CDRs have become the basis for simulating
  epidemics~\cite{Lima_etal13}, quantifying linguistic
  barriers~\cite{Amini_etal13} and optimising public
  transport~\cite{Berlingerio_etal13}.
  Here we apply geospatial techniques to address several questions
  related to social and economic development. 
  How is mobile phone infrastructure related to its use
  (Sec.~\ref{antenna_location} and \ref{spatial_correlation})? 
  How much energy is needed to operate the network (Sec.~\ref{energy})?
  Is the road infrastructure adapted to the population mobility
  patterns (Sec.~\ref{mobility})?
  
  %%%%%%%%%%%%%%%%%%%%%%%%%%%%%%%%%%%%%%%%%%%%%%%%%%%%%%%%%%%%%%% 
  %% Mapping base station locations with respect to population %%
  %% density                                                   %%
  \section{Mapping base station locations with respect to population
    density}
  \label{antenna_location}
  Where to place the base stations that house the antennas is a central
  decision for any mobile communication provider.
  It determines how many people can access the network, the quality of
  calls and the ease with which the provider can operate the facilities.
  Optimising the base station locations is a difficult task, complicated
  by spatially heterogeneous demand and topological obstacles such as
  tall buildings or mountains.
  As a rule of thumb, however, population density is a crucial factor:
  where there are more people, we expect a higher density of base
  stations.
  Conversely, if rural areas with lower population are served by a
  disproportionately low number of base stations, these communities
  would be left with little or no access to the network.
  As mobile communication has enormous potential to improve the lives of
  the rural population (e.g.\ by access to banking and real-time
  information about agricultural commodity prices), one development
  objective must be to provide a roughly equal per-capita number of
  base stations for the entire population of Cote d'Ivoire.
  
  We map the 1238 base station coordinates given in the D4D file 
  \texttt{ANT\_POS.TSV} on a standard latitude-longitude projection
  (left map in Fig.\ 1).
  Since Cote d'Ivoire is close to the equator, such a projection is
  nearly distance-preserving.
  The base stations (coloured dots on the map) are spatially very unevenly
  distributed: in some parts of Abidjan there are more than ten base
  stations per square kilometre, whereas some subprefectures in the
  north of the country have no base station at all.
  That there should be many base stations in Abidjan is quite obvious
  because $\approx 20\%$ of all citizens live in the country's most
  populous city.
  However, whether the number of base stations is proportional to its
  population is not immediately apparent from the latitude-longitude
  projection.
  
  We will thus have to combine the base station coordinates with
  information about the population distribution.
  Here we use census estimates from the AfriPop project
  (http://www.afripop.org)~\cite{Tatem_etal07}.
  Based on these numbers, we project the map of Ivory
  coast so that all regions of the country are represented by an area
  proportional to its population~\cite{GastnerNewman04}.
  Such a density-equalising map -- also known as a cartogram -- has
  become a popular tool to visualise inequality and development
  challenges~\cite{Dorling_etal10}.
  Plotting the base station locations on the cartogram (right map in
  Fig.\ 1) reveals a nuanced picture.
  On one hand, the point distribution is much less aggregated on the
  cartogram and thus is indeed largely proportional to population.
  On the other hand, the points are far from a homogeneous pattern.
  In Abidjan, in particular, a dense cluster of base stations remains
  clearly visible, indicating a disproportionately high per-capita
  connectivity there.
  
  We confirm this observation by calculating the population in the
  base stations' Voronoi cells (Fig.\ 2a).
  (The Voronoi cell of a given base station is the polygon that contains
  the area closer to this base station than to any other.)
  A population-proportional base station distribution would result in an
  equal population inside each Voronoi cell.
  A rank plot of population numbers (Fig.\ 2b), however,
  has a clear S-shape: although most cells have a population of
  around $15\,000$ (mean $15\,500$, median $12\,897$), there are
  outliers in both directions.
  Interestingly, the 16 lowest ranked cells are all in Abidjan, making
  it by far the region with the highest per-capita base station density.
  By contrast, the Voronoi cells with the largest populations are in
  rural areas near inland borders (e.g.\ the second ranked base station at
  $7.267^\circ$~N, $8.160^\circ$~W is 20 km east of the
  Liberian border and the fifth ranked at $9.803^\circ$~N,
  $3.303^\circ$~W is 6 km south of the border with Burkina Faso) or near
  smaller cities (the top and third ranked base station are only a few
  kilometres outside Bouak\'e and the fourth and sixth ranked near
  Korhogo, the country's third and seventh largest cities respectively).
  Because many facility location models suggest that a fair distribution
  of resources should intentionally be skewed in favor of less populated
  areas \cite{GastnerNewman06,Gastner11}, our finding suggests these
  regions as targets for a future expansion of the network. 
  
  %%%%%%%%%%%%%%%%%%%%%%%%%%%%%%%%%%%%%%%%%%%%%%%%%%%%%%%%%%%%%%%% 
  %% Spatial correlation between the population density and the %%
  %% number of calls                                            %%
  \section{Spatial correlation between the population density and the
    number of calls}
  \label{spatial_correlation}
  Recent studies of mobile phone records in developed
  countries~\cite{Schlaepfer_etal12} have argued that the number of human
  interactions in cities increases faster than linearly with the city
  population.
  This poses the question: does the number of calls in Cote d'Ivoire
  depend similarly on population density?
  We count the population and the number of calls on a square grid.
  We investigate squares of size 5~km $\times$ 5~km, 10~km $\times$
  10~km and 20~km $\times$ 20~km.
  We generally find that the number of calls is less correlated for
  smaller than for larger populations so that we divide the data
  into two distinct regions: one for sparsely and another for densely
  populated squares.
  We show the results for ordinary least-squares fits of the form
  $\log(\text{number of calls}) = a\log(\text{population}) + b$ in Fig.\
  3a for the 5~km $\times$ 5~km grid.\footnote{
    Five kilometres is approximately the reception radius of a base
    station, which is the relevant length scale in this problem. However,
    we also state the results for the other grids in table 1.}
  In the densely populated regime (population $>10\,000$), regression
  yields a slope $a=0.87$ with a 95\% confidence interval
  $[0.70,1.03]$.\footnote{Because the logarithm of zero is undefined,
    the regression is calculated by ignoring cells where there were no calls.}
  (The formula for computing confidence intervals can be found for
  example in~\cite{Warton_etal06}).
  For larger sizes (10~km $\times$ 10~km, 20~km $\times$ 20~km) the
  least-squares exponent for dense populations increases, but all 95\%
  confidence intervals include $1$, the dividing line between sub- and
  superlinear scaling (see table 1).
  This finding remains true even if the call intensity is measured
  by the total duration rather than the number of calls.
  Hence, the available data give neither sufficient evidence for nor against
  superlinear scaling for large populations.
  
  For small populations, however, superlinear scaling can be firmly
  ruled out.
  In this regime, the least-squares exponents for the 5~km $\times$ 5~km
  and the 10~km $\times$ 10~km grids are not even significantly
  different from zero, so that population hardly influences the call
  intensity at all.
  The explanation lies in infrastructure located away from population
  centres.
  Among the 5 km $\times$ 5 km squares with a population below $10\,000$,
  three of the ten squares with the largest number of calls are near the
  Buyo hydroelectric plant ($6^\circ14'$~N, $7^\circ3'$~W).
  The other seven squares in the top ten are near major highways (San Pedro
  -- Betia Road, San Pedro -- Tabou Road, A4, A6, A7, A8 and A100).
  These locations are in zones with low population density, but the local
  infrastructure generates a relatively high call intensity.
  
  Despite the weak correlation between calls and population size, the
  spatial distribution of calls is far from random.
  In Fig.\ 3b we plot the spatial autocorrelation function $C_\text{call}$
  on the 5~km $\times$ 5~km grid.
  Although $C_\text{call}$ decays quickly, the correlation is
  nevertheless $>0.1$ up to a distance of $\approx15$~km.
  For comparison, we also plot the autocorrelation $C_\text{pop}$ of the
  population.
  $C_\text{pop}$ is generally a little larger than $C_\text{call}$, but
  it decreases at a similar rate.
  It remains an intriguing question for future research whether both
  autocorrelations are generated by similar social mechanisms.
  In particular, an analysis based on a more careful socio-economic
  definition of ``city size''~\cite{Arcaute_etal13} may still unearth
  more details.
  
  %%%%%%%%%%%%%%%%%%%%%%%%%%%%%%%%%%%%%%%%%%%%%%%%%%%%%%%%%%%%%%%%%% 
  %% Energy and carbon footprint of wireless cellular networks in %%
  %% Cote d'Ivoire                                                %%
  \section{Energy and carbon footprint of wireless cellular networks in
    Cote d'Ivoire}
  \label{energy}
  In this section we estimate the energy and greenhouse gas (GHG)
  emissions, contributing to climate change, of the wireless cellular
  network in Cote d'Ivoire and compare its share of the national
  GHG emissions with that of wireless networks in other countries.  
  
  While mobile network operators are increasingly transparent about
  their environmental impact and GHG emissions, few data are available
  about the energy consumption and resulting GHG emissions of mobile
  networks in developing countries.
  Moreover, the Orange Cote d'Ivoire (OCI) data permits discussing
  energy consumption of parts of the entire network in relationship to
  population density.
  Thus, this work contributes to ongoing research that
  investigates the direct environmental impact of ICT (Information and
  Communication Technology) of systems in general, independent of
  development contexts, by estimating an entire network's footprint from
  the number of its base stations.
  
  Much research has shown that mobile technologies are an
  important instrument of current information and communication
  technologies for development (ICT4D) strategies, for example
  \cite{Aker2010}. 
  On the other hand, the increasing deployment of these technologies can
  result in increasing GHG emissions, sometimes labelled ``footprint'',
  which recently has also received increasing interest by the community
  of ICT4D researchers \cite{Roeth2011,Paul2012}. 
  It is our aim to contribute to a more informed discussion through
  provision of quantitative estimates of energy consumption and GHG
  emissions.
  We want to precede this analysis with a qualification: in or outside
  of a development context the analysis of environmental impact of a
  technical system and its results can stand separately from the
  interpretation of these results towards decision making for policy
  formation.
  In this text we estimate the annual GHG emissions of the mobile
  network in Cote d'Ivoire and suggest directions for existing
  or future development of these networks from the perspective of
  their technical operation. 
  However, this analysis would only provide an incomplete basis for
  policy making towards a development strategy as it does not include an
  analysis of the social or economic impacts and benefits of the
  wireless network.
  
  The goal of our assessment is to estimate the national energy
  consumption and GHG emissions using the number of base
  stations as an input parameter. This requires an estimate of the power
  consumption per base station and the overhead from the remaining parts
  of the network.
  Depending on its type, the power consumption of a base station can
  vary between 800 and 2800~W (estimations presented in \cite{schien}).
  Without additional information about the specific types of base
  stations, the OCI data can only be parameterised with average data. 
  Additionally, an assessment of the energy consumption of a mobile
  network should include all relevant system parts in order to enable
  greater transferability of results. 
  We assume the following composition of the wireless network:
  the base stations, which house the antennas and amplifiers, and
  auxiliary equipment for cooling and power transformation provide the
  radio signal to subscribers. 
  They are controlled by several base station controllers and a few
  mobile service centres to which they are connected via a radio or
  fixed network. 
  This network also provides connectivity with the Internet or networks of
  other operators.
  In our estimate of the GHG emissions we had to make some simplifying
  assumptions about the network infrastructure.
  We estimated the energy consumption for a single base
  station (including overhead for other system parts such as base
  station controllers) of around 2100~W based on similar assumptions made
  in \cite{schien} and \cite{Schien2013} that are based on publicly
  available data by Vodafone.
  This value is a top-down estimate based on the total energy
  consumption of the network and the total number of base
  stations. 
  The corporate responsibility report of the Vodafone Group states that
  in 2011 the company globally operates 224\,000 base stations and that
  the energy consumption was 4117~GWh \cite{Vodafone2011}.
  This value does not account for energy consumption in offices. 
  Given that the average power consumption per base station is around
  1.5~kW, the resulting value of 2100~W per base station is plausible
  and further corroborated by other studies such as \cite{EunsungOh2011}
  who state that the energy consumption of the base stations constitutes
  60-80\% of the total energy consumption of the network.
  
  An estimate of the contribution of the remaining parts of a mobile
  operator's organisation to energy and carbon footprint can, for
  example, be based on corporate social responsibility reports by
  Vodafone and O2, which state that the network accounts for around 80\%
  to 90\%  of an operator's energy consumption
  \cite{O22011,VodafoneDeutschland2011} and constitutes a similar
  portion of its GHG emissions \cite{VodafoneGroup2012}.
  The GSMA Mobile Green Manifesto report \cite{GSMA} makes similar
  assumptions.
  We assume that these ratios also apply to the OCI network and networks
  of other operators in Cote d'Ivoire.
  
  Based on the data inventory we have a precise count of mobile base
  stations (1238). 
  In order to estimate the total annual national energy consumption by
  mobile networks we had to also estimate the number of base stations by
  competitors of OCI in addition to the power consumption by the other
  system parts. 
  We assumed that all mobile operators deploy their network on average
  with similar density. Based on the market share of subscriptions
  (between 33\% OCI \cite{Abidjan.net} and 35\% \cite{LaLiberation}) we
  estimate that the total number of base stations in Cote d'Ivoire is
  around 3700. 
  
  Given these assumptions, we estimate that the GHG emissions of the
  wireless mobile networks by all operators in Cote d'Ivoire amount to
  about 29.1 kilo tonnes carbon dioxide equivalent per year ($ktCO2e$).
  This is about 0.4\% of the total annual carbon emissions of 
  6596.933~ktCO2e \cite{WorldBank2009}.
  We further estimate the energy consumption to be 68~GWh which is about
  1.9\% of the total annual energy production of Cote d'Ivoire
  \cite{MarketEconomics2009}.
  Compared to the pro-rata energy consumption and GHG emissions by
  mobile networks in Germany, this value is relatively high. 
  Based on publicly available data by Vodafone Germany energy
  consumption by the network accumulated circa 600~GWh
  \cite{VodafoneDeutschland2011} in 2011.
  Assuming other wireless networks are equally efficient and Vodafone's
  market share of 32.97\% \cite{Wikipedia2011b}, the energy consumption
  by mobile networks in Germany would be about 0.3\% of the absolute
  energy consumption in the country in 2011
  \cite{StatistischesBundesamt2011}.
  In \cite{GSMA} it is found that on global average, mobile networks
  result in 0.2\% of all GHG emissions.
  Based on the Vodafone data, however, the portion of German mobile
  networks of the national GHG emissions is only around 0.1\%.
  
  Given the lack of data on the power consumption by each base station,
  there is a relatively high uncertainty to the estimate of the total
  annual energy consumption by all networks.
  The estimate of the carbon emissions is further affected by
  uncertainty in the parameter for the carbon intensity of
  electricity.
  In OECD countries, base stations are typically operated with energy
  from the electrical grid.
  In developing countries, however, electrical energy is possibly
  supplied by diesel generators to a significant degree.
  Diesel generators result in a greater carbon intensity per generated
  kWh of electricity (0.788~kgCO2-eq/kWh \cite{Sovacool2008}, as
  compared to 0.426~kgCO2-eq/kWh of the average intensity of grid
  electricity).
  
  In table 2 we evaluate the influence of these parameters
  on the estimates of GHG emissions and energy consumption.
  In scenario I, we consider how the energy consumption and GHG
  emissions would change if the average power consumption per base
  station was reduced by 25\% relative to our base line.
  The resulting average power consumption per base station, including a
  portion for remaining network parts, is 1.58~kW.
  In this case the carbon footprint of the network is 0.33\% and
  slightly closer to the global average value estimated by GSMA in
  \cite{GSMA} of 0.2\%.
  
  Secondly, we evaluate the scenario that half of the electricity
  consumed by the base station was provided by diesel generators and the
  other half by the electrical grid which would increase the carbon
  intensity from 0.426 kgCO2e/kWh to 0.602 kgCO2e/kWh.
  We include the assumption that this would free capacity in the
  electrical grid.
  In our scenario, the mobile network would consume 0.95\% of Cote
  d'Ivoire's electrical energy and constitute 0.62\% of the national GHG
  emissions.
  We also considered evaluating more complex assumptions about the
  types of base stations. Such a scenario would assume that network
  planners generate relatively precise predictions of demand in a
  cell. 
  However, the number of outgoing calls that we plot in Fig.\ 1 together
  with incoming calls are only a possible proxy to overall demand of
  voice traffic.
  Data services and number of calls at peak time must both be considered
  to estimate the minimum capacity of a base station. 
  We believe that the results of such a scenario would have too much
  uncertainty to bring significant value for our discussion.
  
  Given this sensitivity analysis it remains clear, that mobile
  networks in Cote d'Ivoire contribute to a greater degree to the total
  GHG emissions of the country than those in Germany. 
  One of the main reasons for this difference is likely to be the
  contrasting structure of the German and Ivorian economy to which the
  energy intensive manufacturing industry in Germany is likely to
  contribute.
  This assumption is also supported by a comparison of street lighting
  as another energy consuming infrastructure.
  A report by the World Bank mentions in passing that 400\,000 public
  street lights are operated in Cote d'Ivoire \cite{WorldBank2007}.
  Assuming that street lights have a power consumption between 35 and
  400~W \cite{BBCNews} each, they constitute a share of the total energy
  consumption in Cote d'Ivoire between 1.4 and 16 percent.
  In contrast, the street lighting in Germany constitutes only 0.56\% of
  the total energy consumption \cite{BundesministeriumdesInneren}.
  
  Interestingly, if apportioned to each subscriber, the annual energy
  consumption of the OCI mobile network is 3.83~kWh/sub which
  is much lower than the same metric for customers of Vodafone Germany
  (16.5~kWh/sub).
  The value is also a lot lower than the values reported in
  \cite{Malmodin2010} (values between 7 and 34~kWh/sub with an average
  of 16.7~kWh/sub).
  In the case of Cote d'Ivoire, this is likely to be partly the result
  of a sparser deployment of base stations, in particular outside of
  Abidjan as we illustrate in Sec.\ \ref{antenna_location}. 
  One contributing factor to this sparser deployment is likely to be the
  lower degree of urbanisation (52\% compared to 74\% Germany
  \cite{WorldBank2009}). 
  Another factor is the delayed introduction of data services to Cote
  d'Ivoire.
  Third generation services are only just being introduced to this
  market.
  %%%%%%%%%%%%%%%%%%%%%%%%%%%%%%%%
  %%%%%% New: energy consumption by phones and chargers
  Energy consumption on the user side of the mobile network by mobile
  phones and chargers is significantly lower than this. Phone models
  such as the Nokia 108 Dual-Sim \cite{Nokia2014} draw a power between
  as little as 5.86~mW and about 0.254~W while talking. Estimating the
  total energy consumption over the year depends on the user
  behaviour but even if customers spent 1~h talking on the phone per day, which is about 4
  times higher than the average time spent in Europe, the total energy
  consumption by the phone would be about 0.18~kWh which is about
  5\% of the per-user energy consumption by the equipment on the
  operator side. Similarly, the energy consumption by chargers will vary if they remain plugged in
  while not connected to the phone. Yet, even if they are never
  disconnected, their power draw would result in a total energy
  consumption of 0.876~kWh which is about 23\% of the operator side equipment.
  
  These figures have relevance to the ICT4D community because
  development in Cote d'Ivoire can be seen as indicative for many other
  African countries.
  Cote d'Ivoire currently is among the countries with the highest mobile
  phone penetration \cite{GSMA2012}.
  As our estimates illustrate, the uptake of mobile phone technologies
  is accompanied by an increase in energy consumption.
  For Cote d'Ivoire specifically, it is likely that the energy
  consumption by the network will increase in the near future with an
  increasing adoption of data services.

  %%%%%%%%%%%%%%%%%%%%%%%%%%%%%%%%%%%
  %%%% New: Additional energy consumption from expansion to north
  Meanwhile, network coverage in Cote d'Ivoire is not homogenous as was
  described in Sec.~\ref{antenna_location}. The Voronoi plot of the
  country in Fig.\ 2a shows that the Voronoi cells around base
  stations are significantly larger in the region north of the 8th degree
  latitude. Indeed, the average population per cell in this northern
  region is $26\,405$ while it is $14\,414$ in the south. A further 94
  base stations were needed in the north in order to achieve the same
  population density per base station which would increase the energy
  consumption in Orange's network by 7.6\%.

  %%%%%%%%%%%%%%%%%%%%%%%%%%%%%%%%%%%%%%%%%%%%%%%%%%%%%%%% 
  %% Detecting important routes for inter-city mobility %%
  \section{Detecting important routes for inter-city mobility}
  \label{mobility}
  
  CDRs provide a cheap and efficient source of data to study human
  mobility patterns at a large scale \cite{Becker_etal13}. Yet they
  suffer from limitations that need to be carefully considered, and in
  some case dealt with, to ensure the validity of the observations. A
  key limitation is the sparse and heterogeneous sampling of the
  trajectories, as the location is not continuously provided but only
  when the phone engages in a phone call or a text message
  exchange. Moreover, the spatial accuracy of the data is determined
  by the local density of base stations. When estimating mobility from
  CDRs, different approaches have been developed in the literature
  (see Fig.\ 4 for illustration). 
  
  First, researchers interested in statistical models of human
  mobility have adopted a Brownian motion approach
  \cite{Gonzales_etal08}, where each individual is considered as a
  particle randomly moving in its environment. Mobility is considered
  as a path between positions at successive position
  measurements. Authors have observed statistical properties
  reminiscent of Levy flights, together with a high degree of
  regularity. Yet, the usefulness of these observations is limited by
  the bursty nature of phone activity, as burstiness is expected to
  alter basic statistical properties of the jumps, such as their
  distance distribution (see Fig.\ 5). Even in studies where
  the positions are evaluated at regular intervals, the nature of the
  jumps remains unclear, as the method tends to detect short trips due
  to localisation errors, and is blind to the type of the places
  sampled from the real trajectory. As a side note, let us mention
  recent work using geo-localised web services, such as Foursquare,
  where users voluntarily check-in at places
  \cite{Noulas_etal11,Noulas_etal12}. Foursquare check-ins are also
  characterised by a bursty behaviour, but they provide a GPS
  accuracy, and semantic information (at the office, travelling, etc.)
  that might solve the aforementioned problems.
  
  The second approach relies on the idea that mobility consists of
  moving from one place to another. The observation of mobility
  patterns thus requires one to define and identify important
  locations. A trajectory is seen as a set of consecutive locations
  visited by the user. Important locations can either be defined as a
  place where a user spends a significant amount of time, which he
  visits frequently, or where he has stopped for a sufficiently long
  time
  \cite{Becker_etal13,Becker_etal11,Isaacman_etal10,Quercia_etal11}. This
  approach provides a more intuitive picture of mobility, where the
  sampling is determined by the periods of rest of the user. However,
  it is blind to the multi-scale nature of human mobility, as it
  requires the parametrisation of thresholds in time and in space to
  identify important locations. The value of the threshold and the
  corresponding granularity of the places depends on the system under
  scrutiny, say cities for international mobility or rooms for human
  mobility inside hospitals \cite{Lucet_etal12}. 
  
  When measuring human mobility from CDRs, it is important to remember
  that mobility is about space and time. Both aspects must be
  carefully considered to provide a faithful description of human
  trajectories, especially in situations where the sampling of the
  data is heterogeneous. For this reason, each transition should be
  remembered as a jump in space over an interval in time and, if
  possible, be put in relation to the previous and following
  transitions. Contrary to the universality viewpoint of
  \cite{Gonzales_etal08}, not all transitions are alike. On the
  contrary, it is possible to extract different information and
  different types of mobility patterns by focusing on different
  regions in space-time. This filtering has been adopted in various
  studies, but usually either in space or in time. Let us mention
  \cite{Quercia_etal11}, where transitions between identified places
  are considered only if they are registered within two hours of each
  other; in \cite{Isaacman_etal10} the daily range of mobility is
  calculated, and in \cite{Wang_etal12} a trip is defined as a
  displacement between two distinct base stations occurring within one
  hour in each time period. More complex filters can be defined on
  so-called {\it handoff patterns}, that is a sequences of cell towers
  that a moving phone uses while engaged in one voice call, e.g.\ in
  \cite{Becker_etal11} where only sequences of more than 5 cell towers
  are included. Let us note that a filtering in space and in time
  allows for the selection of a characteristic velocity and, if
  needed, of the removal of noisy transitions occurring at a small
  spatial scale, e.g.\ transitions between neighbouring cells of a
  static user, or long temporal scale, e.g.\ transitions over several
  days where several intermediate steps are expected to be missing.
  
  This overview of recent research suggests direct applications that
  would be of particular interest in a developing country, where
  empirical data on human mobility tend to be lacking. Using the
  aforementioned methodologies, it would be possible, for instance, to
  identify and map nationwide commuting patterns. Traffic tracking and
  route classification would also be possible after additional data is
  collected from test drives or signal strength data collected by
  high-resolution scanners \cite{Becker_etal11}.
  In this work, we illustrate the potential benefits of a CDR analysis
  by focusing on the detection of high-traffic roads between
  cities. Such a detection might help deploy new infrastructure where
  the population actually needs it, e.g. in regions where mobility is
  high but the infrastructure is poor. Finding high-traffic roads
  requires one to filter transitions in the two-dimensional space of
  Fig.\ 5. To do so, we apply the following procedure. We
  consider only transitions in a certain velocity range and occurring
  in less than a predefined time interval. Our choice of velocity
  range for car mobility is $[15,150]{\rm km/h}$, in order to discard
  pedestrian motion and noisy points, i.e. due to antenna switching
  instability. For our analysis, we have used the data from
  \texttt{POS\_SAMPLE\_X.TSV} source files, containing separate users'
  traces, in the form of a list of user - antenna - timestamp for
  each call or SMS, together with the antenna positions
  from the \texttt{ANT\_POS.TSV} file. The lower bound for the time
  interval between two points has been set as the minimal value between
  two actions. For the upper bound, a one hour limit has been chosen
  in order to balance between sufficient data points and accuracy.
  Moreover, to remove noisy connections and to identify persistent
  motion, we have removed weak transitions between antennas,
  i.e. occurring less than 10 times. This operation leads to a
  fragmentation of the network into connected components which we
  further exploit by keeping only components composed of at least ten
  vertices. This operation has the advantage of removing undesired
  connections due to antenna switching and not associated to motion.
  Our results are robust under variations of the above parameters.
  
  The described technique gives a good approximation of the most
  important human migration pathways (see Fig.\ 6) and thus can be used for
  alternative road construction or improvement. Interestingly, it also
  allowed us to identify unknown roads, which we could validate {\it a
    posteriori}. Examples are shown in Fig.\ 7 and Fig.\ 8 where roads
  that were absent in Microsoft maps but found by our algorithm are
  found in maps  provided by OpenStreetMap and Yahoo
  respectively. This can be of particular interest for a
  semi-automated map improvement technique: if a strong connection is
  found from the CDR information, but there is no road on the map, it
  should be analysed carefully whether an existing road has so far been
  overlooked.
  
  %%%%%%%%%%%%%%%% 
  %% Conclusion %%
  \section{Conclusion}
  In this article we have presented how an analysis of the Orange D4D
  mobile phone data base reveals important patterns of communication
  infrastructure and mobile phone use in Cote d'Ivoire.
  The placement of base stations is biased towards Abidjan so that
  one development goal is an enhancement of the network in smaller
  cities and rural regions.
  We estimate that the network currently consumes between 2.88 and
  3.83~kWh of energy annually per subscriber.
  Although this figure is less than in an industrial country such as Germany, the
  fraction of the national energy consumption spent on mobile
  telephony (estimated between 0.95\% and 1.90\%) is actually higher. 
  Finally, we argued that mobility data from CDRs need further
  filtering to extract truly meaningful commuting patterns.
  We used the mobility traces that were part of the Orange D4D database
  to demonstrate how the main roads in Cote d'Ivoire can be identified.
  
  %%%%%%%%%%%%%%%%%%%%%%%%% 
  %% Competing Interests %%
  \section*{Competing interests}
  The authors declare that they have no competing interests.
  
  %%%%%%%%%%%%%%%%%%%%%%%%%%%% 
  %% Authors' Contributions %%
  \section*{Authors' contributions}
  MTG performed the analysis of base station locations. HY and MTG
  calculated the correlation between phone calls and population
  density and drafted the text of sections 2 and 3. DS calculated the
  energy consumption and drafted section 4. VS performed the analysis of
  mobility traces. VS and RL drafted section 5. All authors reviewed and
  approved the complete manuscript.
  
  %%%%%%%%%%%%%%%%%%%%%% 
  %% Acknowledgements %%
  \section*{Acknowledgements}
  \ifthenelse{\boolean{publ}}{\small}{}
  We thank Orange for making the D4D data set available.
  VS and RL acknowledge financial support from FNRS.
  MTG is grateful for financial support from the University of Bristol
  and the EPSRC Building Global Engagements in Research (BGER)
  grant. He also acknowledges support from the European Commission
  (project number FP7-PEOPLE-2012-IEF 6-456412013).
  This paper presents research results of the Belgian Network DYSCO
  (Dynamical Systems, Control, and Optimization), funded by the
  Interuniversity Attraction Poles Programme, initiated by the Belgian
  State, Science Policy Office.
  HY acknowledges the support by grants from the Rockefeller Foundation
  and the James McDonnell Foundation (no. 220020195).
  
  %%%%%%%%%%%%%%%%%%%%%%%%%%%%%%%%%%%%%%%%%%%%
  %% The Bibliography                       %%
  %%%%%%%%%%%%%%%%%%%%%%%%%%%%%%%%%%%%%%%%%%%% 

%%%%%%%%%%% 

\ifthenelse{\boolean{publ}}{\end{multicols}}{}

%%%%%%%%%%%%%%%%%%%%%%%%%%%%%%%%%%% 
%%                               %% 
%% Figures                       %%
%%                               %% 
%% NB: this is for captions and  %%
%% Titles. All graphics must be  %%
%% submitted separately and NOT  %%
%% included in the Tex document  %%
%%                               %% 
%%%%%%%%%%%%%%%%%%%%%%%%%%%%%%%%%%% 

%% 
%% Do not use \listoffigures as most will included as separate files

\section*{Figures}
\subsection*{Figure 1 - Base station locations}
\includegraphics[width=\linewidth]{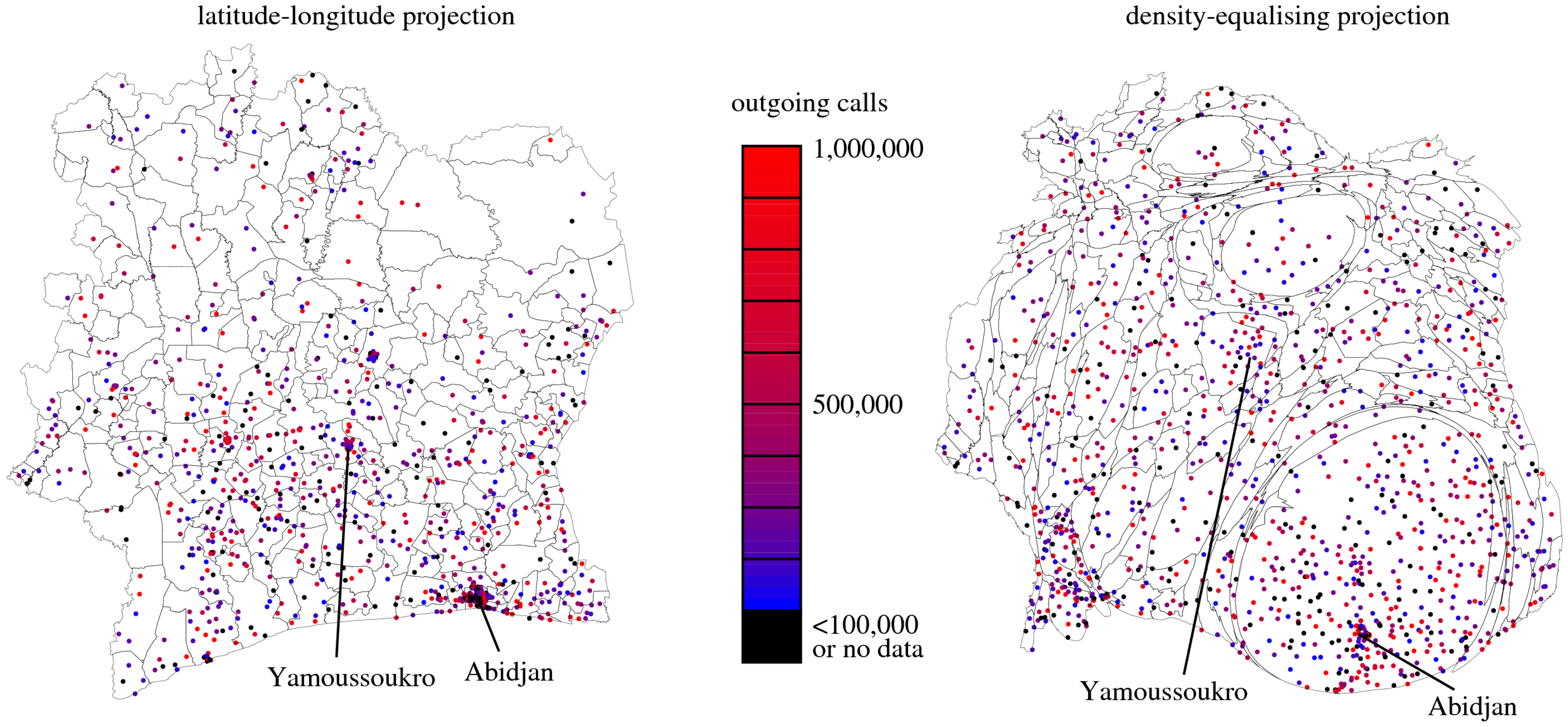}

Base station locations on a conventional longitude-latitude
projection (left) and a cartogram where areas are rescaled to be
proportional to the number of inhabitants (right). 
The colours of the dots indicate the number of outgoing calls at
each base station.
The boundaries of subprefectures are shown for ease of orientation.
The geographic distribution of the base stations are largely
explained by the heterogeneous population distribution so that the
point pattern appears less clustered on the right than on the left.
Still, regions of significantly higher per-capita base station density
remain (see Fig.\ 2), especially in Abidjan, where
even on the cartogram the dots are noticeably aggregated. 
The colours of the dots do not exhibit any clearly visible
large-scale trends.
However, a more careful statistical analysis shows that a
significant correlation between traffic at nearby base stations
exists (see Fig.\ 3b).

\subsection*{Figure 2 - Rank plot of populations in Voronoi cells}
\includegraphics[width=\linewidth]{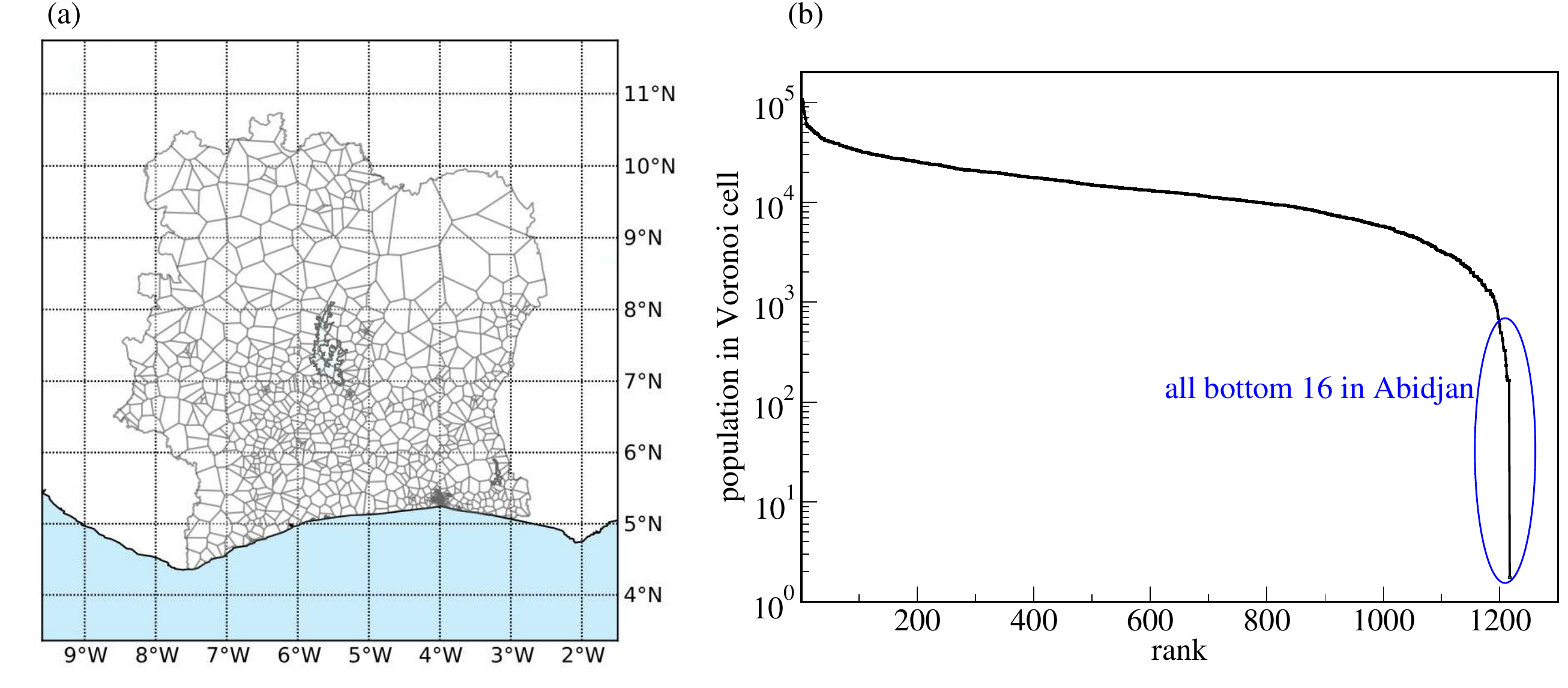}

(a) Voronoi cells of the $1217$ distinct base station locations in the
Orange D4D challenge data base.
(b) Rank plot of the population inside the Voronoi cells.
Although a majority of $616$ cells are within $50\%$ of the median
population ($12\,897$ inhabitants), there are significant outliers at
both top and bottom ranks. While the bottom ranked cells are
predominantly in Abidjan, the top ranks are in rural areas as well
as smaller cities.

\subsection*{Figure 3 - Correlation between population and phone
  call intensity}
\includegraphics[width=8cm]{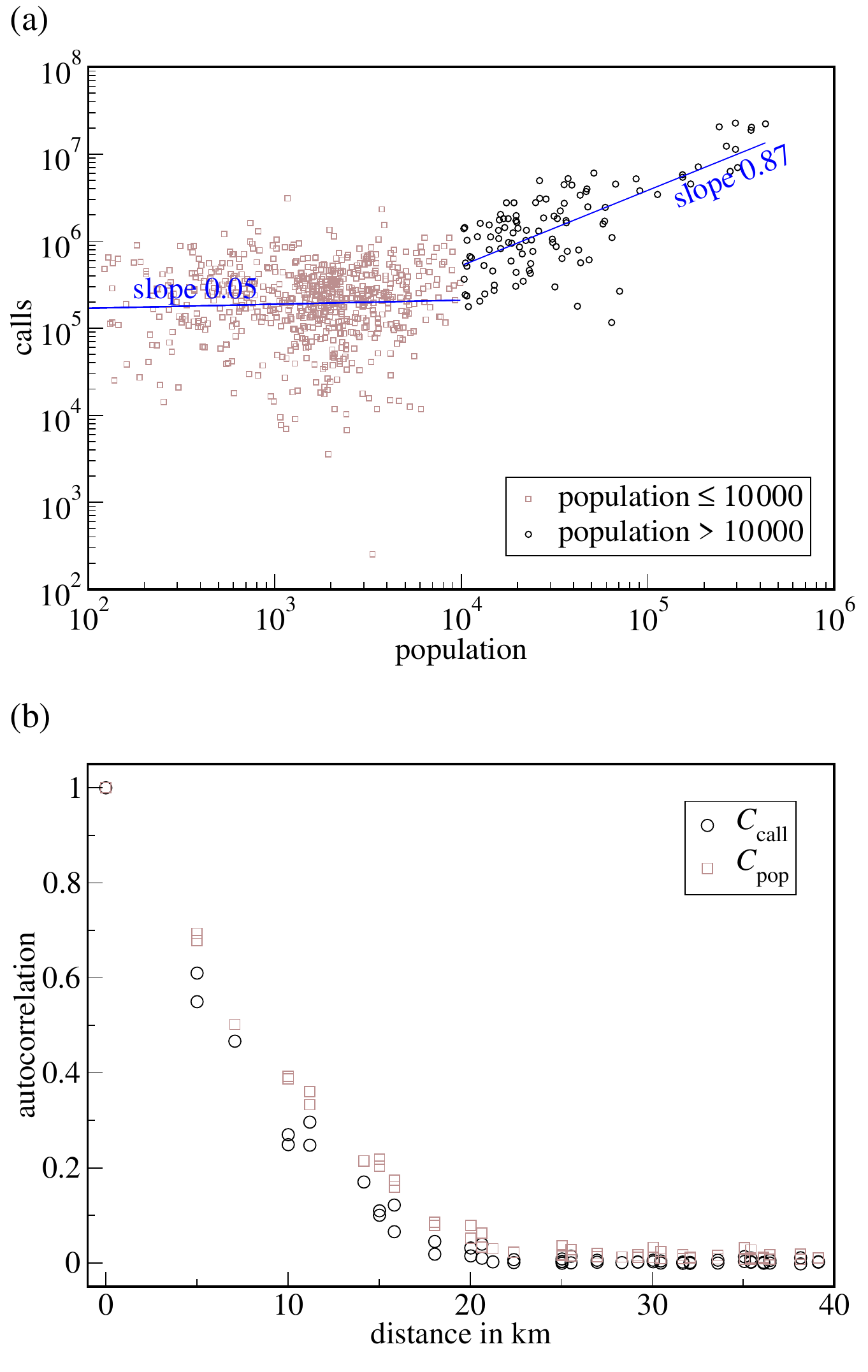}

(a) Scatter plot of the total number of outgoing calls versus the
population.
Both variables are counted on a 5~km $\times$ 5~km square grid.
An ordinary least-squares fit yields different slopes for small
and large populations.
(b) The spatial autocorrelation functions for the number of calls
and the population size exhibit significant non-zero correlations up
to $\approx15$~km.

\subsection*{Figure 4 - Detecting mobility patterns}
\includegraphics[width=\linewidth]{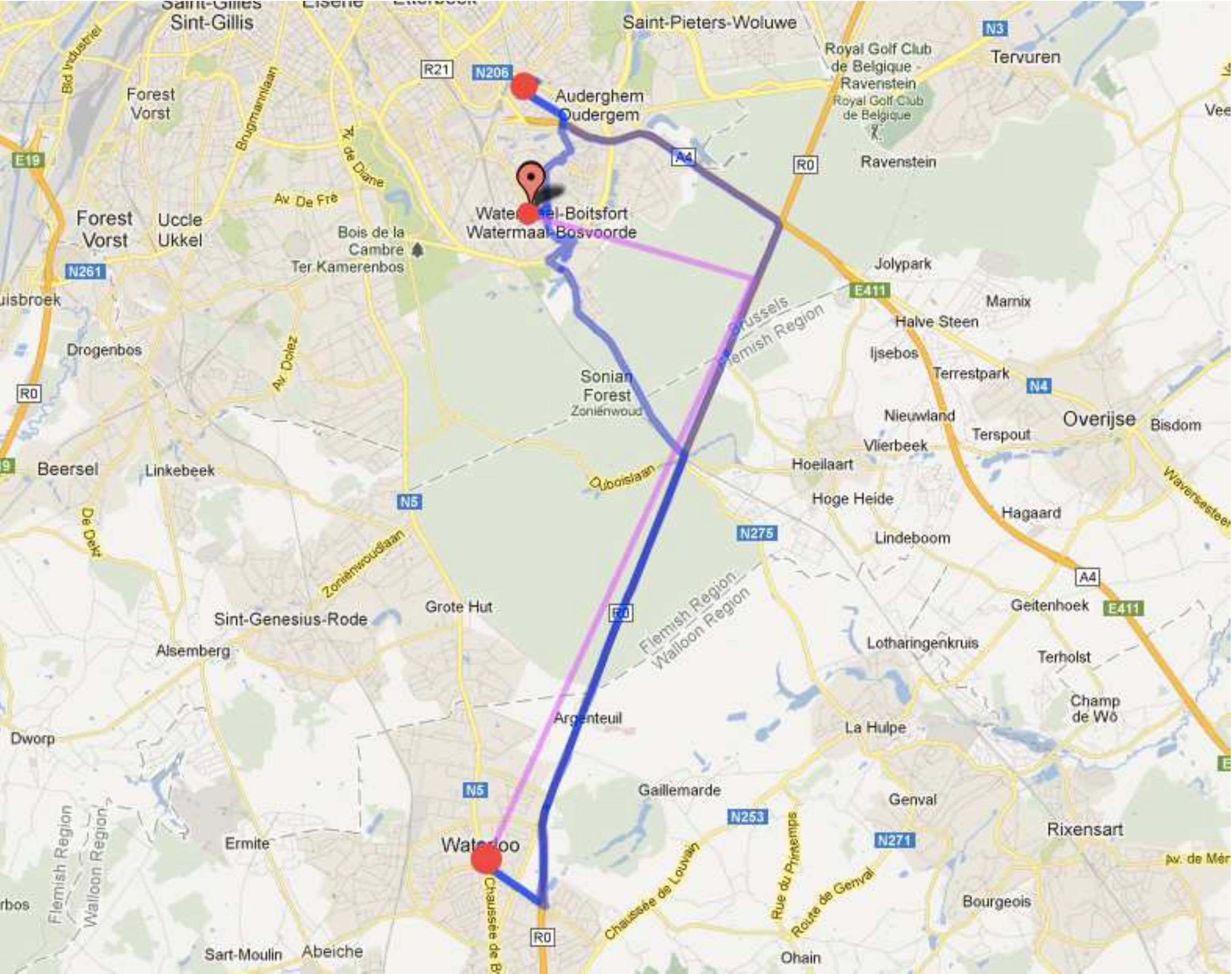}

To illustrate the different ways to uncover mobility patterns from
CDRs, let us focus on the motion of an individual in Brussels, as
measured by his GPS. 
The user took his car in Watermael and went to two shops, one in
Auderghem and one in Waterloo.
The three locations are plotted in red.
Three phone calls were made.
One at home, one on the highway, and one in Waterloo. An approach
where a path is composed of successive position measurements is
shown in pink.
In contrast, an approach where paths are based on important
locations would detect the stop in Waterloo, rightly discard the one
on the highway, but would still be blind to the location in Auderghem.

\subsection*{Figure 5 - Traveled distance versus travel time}
\includegraphics[width=\linewidth]{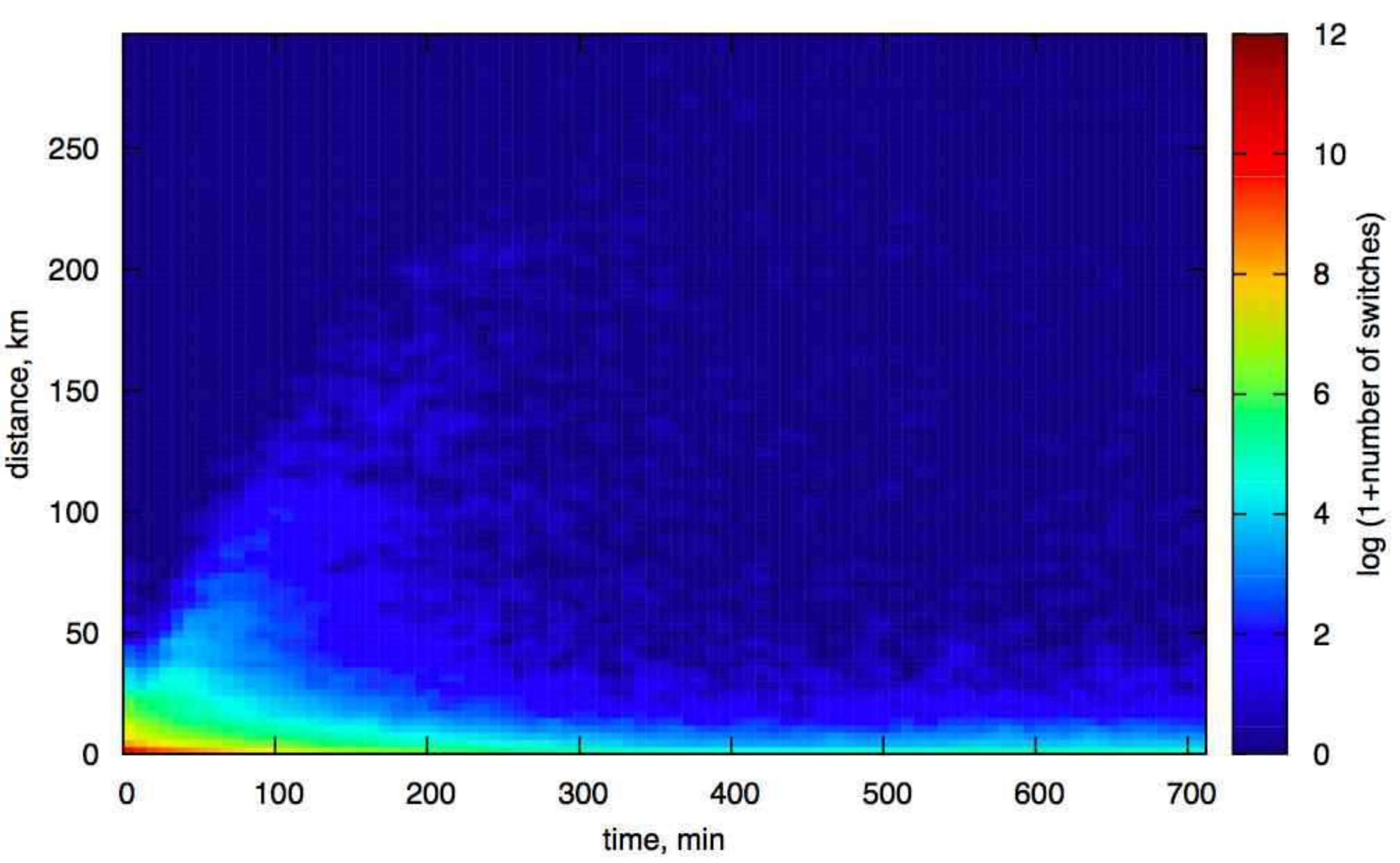}

Heat map of the distances and time intervals between consecutive
CDRs.
The burstiness of phone activity leads to a broad distribution over
time.
Keeping transitions from different regions in the two-dimensional
space allows for the identification of different aspects of human
mobility.

\subsection*{Figure 6 - High-traffic roads from mobile phone records}
\includegraphics[width=0.9
\linewidth]{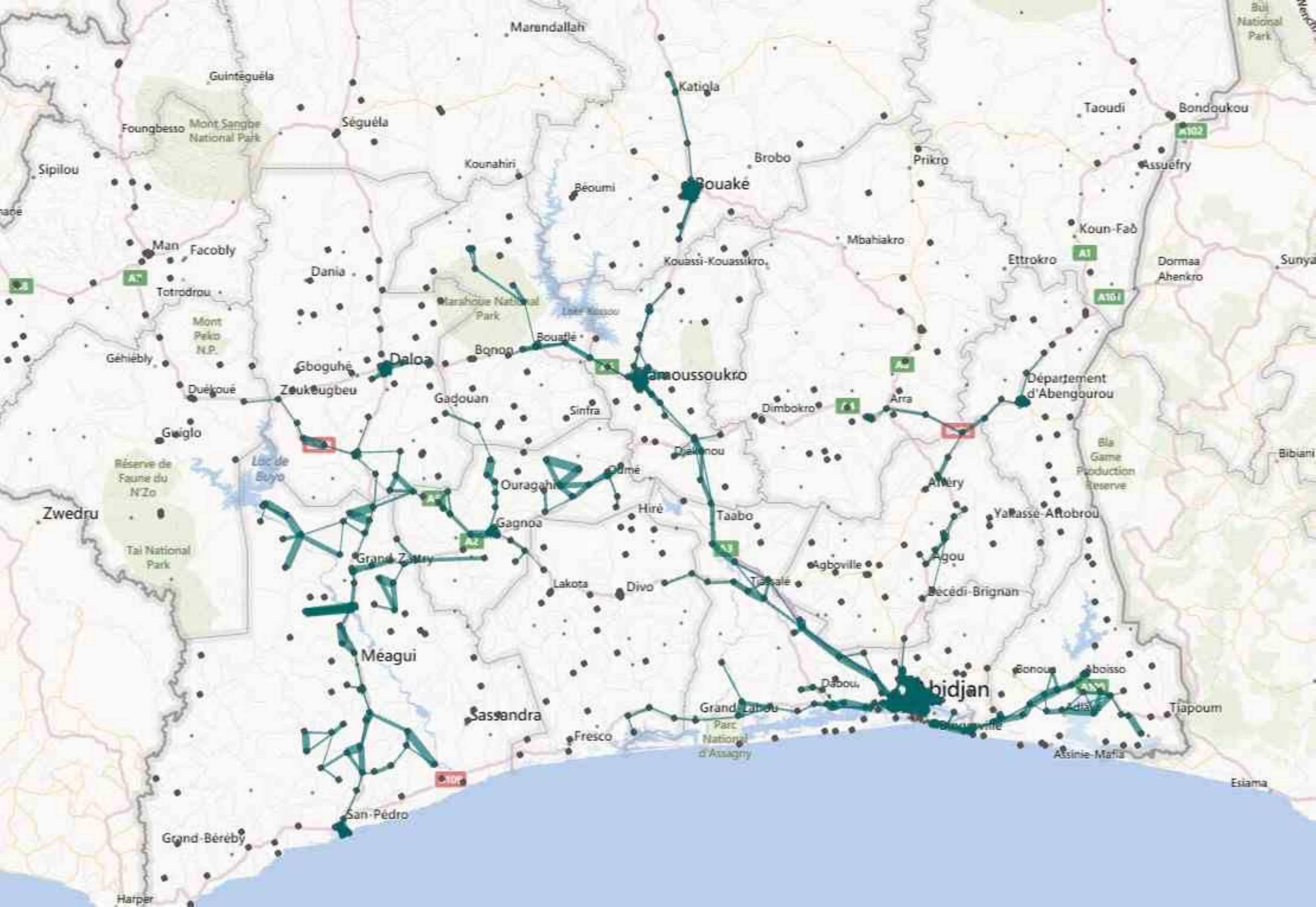}

High-traffic road detection, as obtained from CDR data.

\subsection*{Figure 7 - Road identification on Microsoft and OpenStreetMap maps}

\includegraphics[width=0.9
\linewidth]{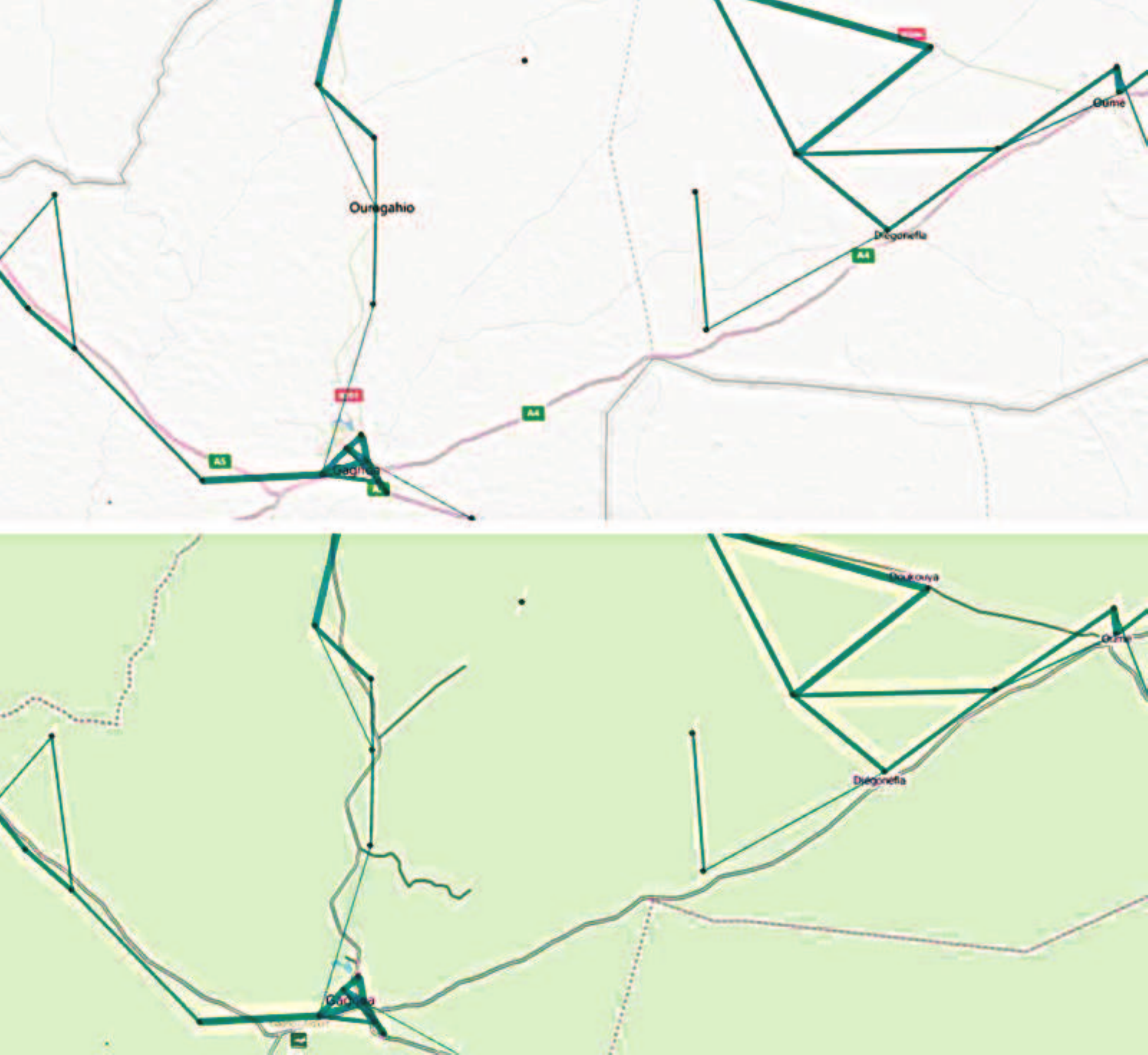}

Roads that are identified from CDR data and are absent on a
Microsoft map (upper panel) can be identified on an OpenStreetMap
(lower panel).

\subsection*{Figure 8 - Road identification on Microsoft and Yahoo maps}

\includegraphics[width=0.9
\linewidth]{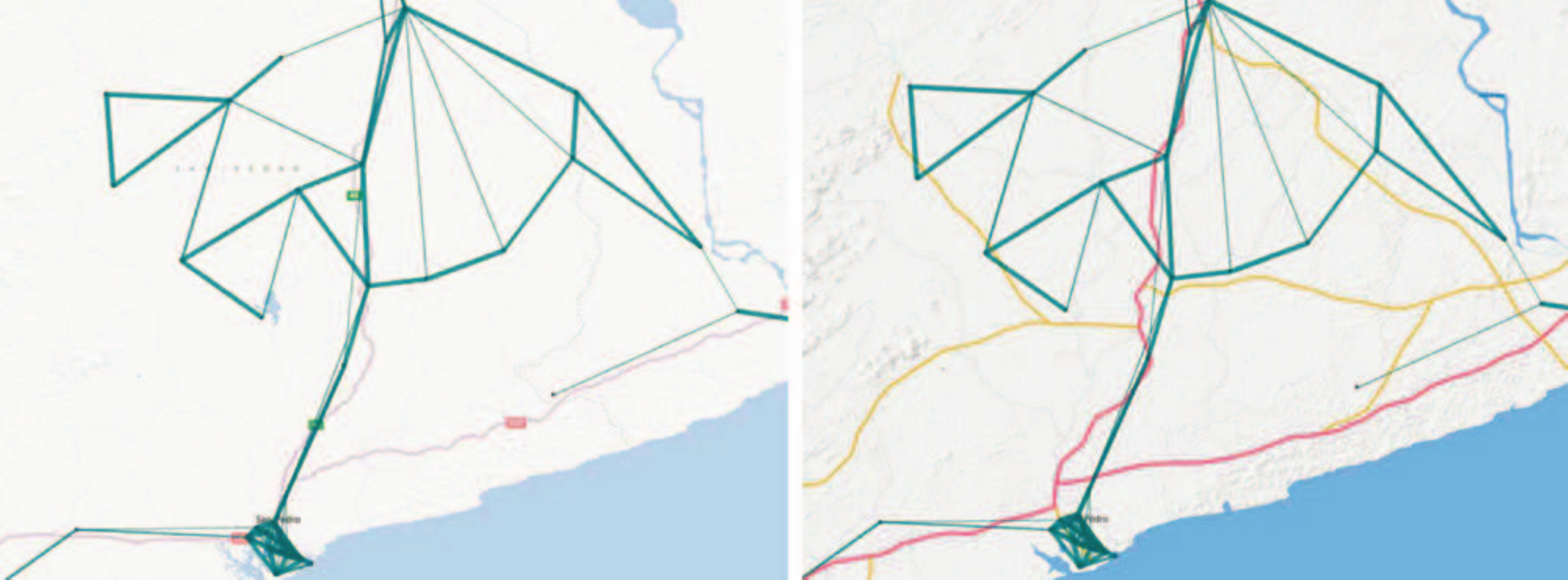}

Roads that are identified from CDR data and are absent on a
Microsoft map (left panel) can be identified on a Yahoo map (right
panel).

%%%%%%%%%%%%%%%%%%%%%%%%%%%%%%%%%%% 
%%                               %% 
%% Tables                        %%
%%                               %% 
%%%%%%%%%%%%%%%%%%%%%%%%%%%%%%%%%%% 

%% Use of \listoftables is discouraged.
%% 
\section*{Tables}
\subsection*{Table 1 - Regression of call intensity versus
  population}
Results of regression to
$\log(\text{intensity})=a\log(\text{population})+b$.
The cutoff between small and large populations is $10\,000$
for the 5~km $\times$ 5~km grid, $20\,000$ for the 10~km $\times$
10~km grid and $40\,000$ for the 20~km $\times$ 20~km grid.
The cutoffs were chosen so that the number of data points in the large
population regime are approximately equal ($105$, $74$ and $80$
respectively).
\par \mbox{}
\par
\mbox{
  \begin{tabular}{c|c|c|c|c|c}
    \hline\hline
    && \multicolumn{2}{|c|}{small population} &
    \multicolumn{2}{|c}{large population}\\
    \hline
    grid & intensity & $a$ & 95\% CI & $a$ & 95\% CI\\
    \hline
    \multirow{2}{*}{5 km $\times$ 5 km} & number of calls & $0.05$ &
    $[-0.04,0.13]$ & $0.87$ & $[0.70,1.03]$\\
    & duration of calls & $0.06$ & $[-0.02,0.14]$ & $0.86$ &
    $[0.70,1.02]$\\
    \hline
    \multirow{2}{*}{10 km $\times$ 10 km}& number of calls & $0.06$
    & $[-0.05,0.17]$ & $1.00$ & $[0.77,1.23]$\\
    & duration of calls & $0.08$ & $[-0.03,0.18]$ & $1.01$ &
    $[0.78,1.23]$\\
    \hline
    \multirow{2}{*}{20 km $\times$ 20 km}& number of calls & $0.27$
    & $[0.11,0.43]$ & $1.24$ & $[0.91,1.57]$\\
    & duration of calls & $0.28$ & $[0.12,0.43]$ & $1.27$ & $[0.93,1.60]$\\
    \hline\hline\hline
  \end{tabular}
}

\subsection*{Table 2 - Energy consumption of Cote d'Ivoire's mobile
  phone network}
Scenarios of alternative input parameters. Scenario 'base'
assumes average values for all parameters, scenario 'I'
assumes a value for the power consumption per base station
compared to 25\% from base. Scenario 'II' assumes that 50\% of the
electricity consumed by base stations is supplied by diesel
generators.\par \mbox{}
\par
\mbox{
  \begin{tabular}{p{.5\columnwidth}|p{.1\columnwidth}|
      p{.1\columnwidth}|p{.1\columnwidth}}
    \hline \hline
    & BASE & I & II \\ \hline
    Carbon intensity of electricity (kgCO2e/kWh) & 0.43 & 0.43 &
    0.60 \\ \hline
    Average aggregate power consumption per BS (kW) & 2.10 & 1.58 &
    2.10 \\ \hline 
    Total national energy consumption by mobile networks(GWh) &
    68.32 & 51.24 & 68.32 \\ \hline 
    National energy consumption by mobile networks (percent of
    total)  & 1.90 & 1.43 & 0.95 \\ \hline 
    Total national GHG emissions by mobile networks (ktCO2e) & 29.11
    & 21.83 & 41.13 \\ \hline 
    National GHG emissions by mobile networks (percent of total)  &
    0.44 & 0.33 & 0.62 \\ \hline 
    Annual energy consumption per subscriber (kWh/sub) & 3.83 & 2.88
    & 3.83 \\ \hline 
    \hline \hline
  \end{tabular}
}

\end{bmcformat}
\end{document}